\documentclass[12pt,preprint]{emulateapj}

\def\be{\begin{equation}}
\def\ee{\end{equation}}
\def\bea{\begin{eqnarray}}
\def\eea{\end{eqnarray}}

\usepackage{graphicx}

\begin{document}

\title{XRF 100316D/SN 2010bh: clue to the diverse origin of nearby supernova-associated GRBs}

\author{Yi-Zhong Fan$^{1,2}$, Bib-Bin Zhang$^{2}$, Dong Xu$^{3}$, En-Wei Liang$^{4,2}$ and Bing Zhang$^{2}$}
\affil{$^1$ {Purple Mountain Observatory, Chinese Academy of Science,
Nanjing, 210008, China.}\\
$^{2}$ {Department of Physics and Astronomy,
University of Nevada Las Vegas, Las Vegas, NV 89154, USA.}\\
$^{3}$ {Benoziyo Center for Astrophysics, Faculty of Physics, The Weizmann Institute of Science, Rehovot 76100, Israel.} \\
$^{4}$ Department of Physics, Guangxi University, Guangxi 530004,
China.} \email{yzfan@pmo.ac.cn (YZF), zhang@physics.unlv.edu(BZ)}

\begin{abstract}
X-ray Flash (XRF) 100316D, a nearby super-long under-luminous burst
with a peak energy $E_{\rm p}\sim 20$ keV, was detected by {\it
Swift} and was found to be associated with an energetic supernova SN 2010bh.
Both the spectral and the temporal behavior are rather similar to XRF
060218, except that the latter was associated with a ``less energetic" SN
2006aj, and had a prominent soft thermal emission component in the spectrum.
We analyze the spectral and temporal properties of this burst, and
interpret the prompt gamma-ray emission and the early X-ray plateau
emission as synchrotron emission from a dissipating Poynting flux
dominated outflow, probably powered by a magnetar with a spin period
of $P \sim 10$ ms and the polar cap magnetic field $B_{\rm p} \sim 3\times
10^{15}$ G.
The energetic supernova SN 2010bh associated with this burst
is however difficult
to interpret within the slow magnetar model, and we suspect that
the nascent magnetar may spin much faster with an initial rotation
period $\sim 1$ ms.
It suggests a delay between the core collapse and the emergence
of the relativistic magnetar wind from the star. The diverse behaviors
of low-luminosity GRBs and their associated SNe may be understood
within a unified picture that invokes different initial powers
of the central engine and different delay times between the core
collapse and the emergence of the relativistic jet from the star.
\end{abstract}

\keywords{Gamma rays: general ---Radiation mechanisms:
non-thermal---X-rays: general}

\setlength{\parindent}{.25in}

\section{INTRODUCTION}
After four years of waiting since the detection of X-Ray Flash (XRF)
060218/SN 2006aj, another pair of low-luminosity (LL) GRB - supernova
(SN) association, XRF 100316D/SN 2010bh at redshift $z=0.059$
\citep{Vergani10}, was captured by {\em Swift}
\citep{Gehrels04} on March 16, 2010
\citep{Stamatikos10,Wiersema10,Chornock10,Rau10}, with a detection
rate fully consistent with the population studies of these nearby LL-GRB
events \citep{Coward05,Soderberg06,Liang07,Guetta07}. Before this
event, four
pairs of nearby ($z<0.2$)
secure GRB(XRF)-SN associations
have been identified. These are
GRB 980425/SN 1998bw at $z=0.0085$ \citep[e.g.,][]{Galama98},
GRB 030329/SN 2003dh at $z=0.168$ \citep[e.g.,][]{Hjorth03},
GRB 031203/SN 2003lw at $z=0.105$ \citep[e.g.,][]{Malesani04}, and
XRF 060218/SN 2006aj at $z=0.0331$ \citep[e.g.,][]{Campana06}.
The nature of the GRB/SN connection and the interplay between the GRB
and the SN components are still poorly understood.
In this paper, we analyze and interpret the {\em Swift} BAT and XRT
data of XRF 100316D, paying special attention to the similarities
and differences between the XRF 100316D/SN 2010bh and XRF 060218/SN 2006aj.

\section{The $\gamma-$ray and X-ray observations}
GRB 100316D triggered {\em Swift}/BAT at 12:44:50 UT ($T_{\rm trig}$),
and the XRT began observing the field 137.7 seconds after the BAT
trigger (Stamatikos et al. 2010). A bright, steady un-catalogued X-ray
source was detected. By analyzing the BAT survey data before and after
the trigger time, it is found that this event possibly started at
$\sim (T_{\rm trig} - 1500)$ seconds \citep{Stamatikos10}. However,
the gamma-ray flux of the
source kept almost constant from $T_{\rm trig}-1500$ to $T_{\rm
trig}-500$, before increasing significantly at $T_{\rm
trig}-500$. We therefore take the starting time of this event as
$T_{\rm trig}-500$ seconds.
The host galaxy redshift is $z=0.059$ \citep{Vergani10}.
An associated supernova, SN 2010bh, was detected by Wiersema et al.
(2010) and spectroscopically confirmed by Chornock et al. (2010).

\begin{figure}
 \includegraphics[angle=270,width=0.4\textwidth]{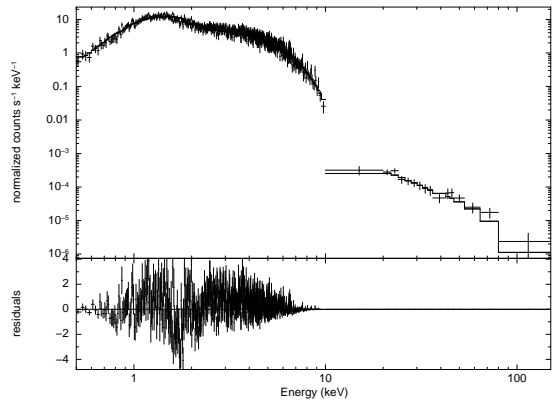}
\caption{The joint BAT-XRT spectrum of XRF 100316D with a Cutoff
Power-law model fit (line) for the time-integrated spectrum from
$T_{\rm trig}+138$ s to $T_{\rm trig}+736$ s. The model
parameters are $\Gamma= 1.32\pm 0.03$ and
$E_{\rm p} = 19.6^{+3.3}_{-2.8}~{\rm keV}$. }
\label{fig:BAT-XRT}
\end{figure}

\begin{figure}
 \includegraphics[angle=0,width=0.45\textwidth]{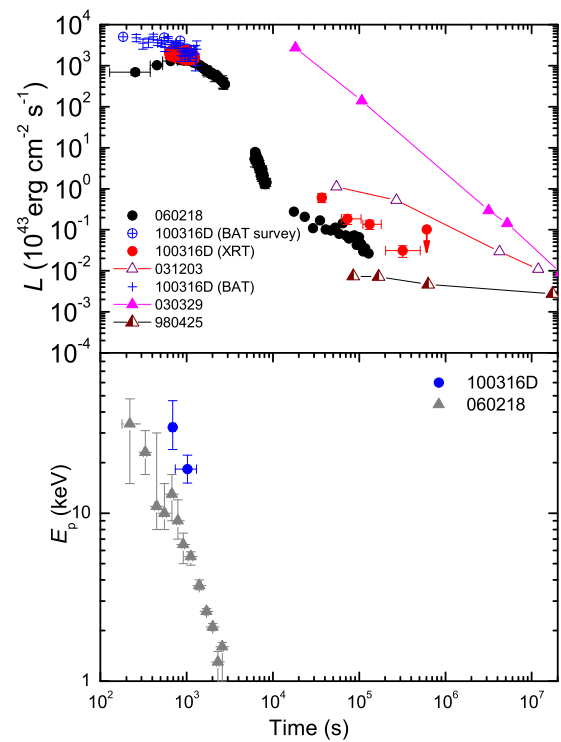}
\caption{{\em Upper:} Unabsorbed luminosity lightcurve in the XRT band of GRB
100316D in comparison with GRBs 980425, 030329, 031203, and 060128. The BAT
data of GRB 100316D are extrapolated to the XRT band. {\em Lower:} Comparison
of the $E_{\rm p}$ temporal evolution of GRB0100316D with GRB 060218. The
$E_{\rm p}$ data of GRB 060218 are taken from Toma et al. (2007)}
\label{fig:XRT}
\end{figure}

We extract the observed lightcurves and spectra of XRF 100316D from
the BAT and XRT event data. The details of our data reduction are
presented in \citet{ZLZ07} and \citet{Liang07a}. The joint BAT/XRT
spectrum from $T_{\rm trig}+138$ s to $T_{\rm trig}+736$ s is well fit
with a Cutoff Power-law model, with a power-law index of $1.32\pm
0.03$ and a spectral peak energy $E_{\rm p}$ of $19.6^{+3.3}_{-2.8}
~{\rm keV}$ ($\chi^2/{\rm dof}=1320/1057$),
Fig.\ref{fig:BAT-XRT})\footnote{The time-integrated
BAT spectrum can be reasonably fitted by a single power $F_\nu \propto
\nu^{-2.5\pm 0.3}$. Together with the
joint BAT-XRT fit, we are able to constrain $E_{\rm p}\sim 20$
keV and therefore define this burst as an XRF.}. This suggests that
the X-ray and the gamma-ray emissions are from the same emission
component.
A time-resolved spectral analysis during the period from $T_{\rm trig}+138$ s
to $T_{\rm trig}+736$ s shows that $E_{\rm p}$ clearly evolves with time, from
$32.6^{+ 14.2}_{-8.5}$ keV (from $T_{\rm trig}+138$ s to $T_{\rm trig}+240$ s)
to $18.3^{+ 3.9}_{-3.2}$ keV (from $T_{\rm trig}+240$ s to $T_{\rm trig}+734$
s). The un-absorbed lightcurves and the temporal evolution of $E_{\rm p}$ are
shown in Figure \ref{fig:XRT}, along with the data of GRB 060218. The late
X-ray lightcurves of the other nearby GRBs, including GRB 980425, 031203, and
030929, are also displayed in Figure \ref{fig:XRT}. The X-ray light curves of
XRF 100316D and XRF 060218 are rather similar. Both have an plateau phase
extending to more than 1000 s. In the case of XRF 100316D, there is an
observational gap between $T_{\rm trig}+736$ s and $T_{\rm trig}+36437$ s. The
late-time observational data (after $T_{\rm trig}+36437$ s) is soft and
consistent with a single power law decay with a decay index $\alpha = 1.31 \pm
0.21$. This is similar to the late X-ray lightcurve of XRF 060218. XRF 060218
also shows a steep decay phase between the plateau and the late single power
law decay phase. Although this is not observed in XRF 100316D, the data before
$T_{\rm trig}+734$ s and after $T_{\rm trig}+36437$ s are consistent with
having such a steep decay segment in between.

\begin{figure}
 \includegraphics[angle=0,width=0.4\textwidth]{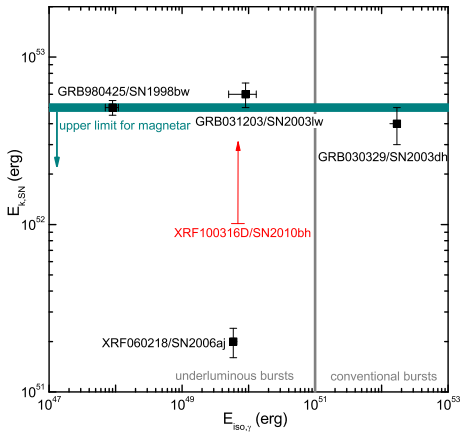}
\caption{The isotropic energy of the prompt emission vs. the
kinetic energy of the supernova outflow.
The kinetic energy of SN 2010bh is estimated to be larger than $\sim 10^{52}$ erg
(see footnote \ref{foot:2}). Other data are taken from
\citet{Li06}. The possible maximum energy $\sim 5\times 10^{52}$ erg
that can be provided by a pulsar with $P\lesssim 1$ ms and $I \sim 2\times 10^{45}~{\rm g~cm^{2}}$ is also plotted.
}
\label{fig:SN-GRB}
\end{figure}

There are however some differences between the two events. First, the X-ray
data of XRF 060218 demand a soft thermal emission component with an evolving
temperature, which has been interpreted as due to shock breakout from the
progenitor star \citep{Campana06}. For XRF 100316D, we find that the data do
not demand such a component. We noticed that \cite{Starling10} claimed a
thermal component with $k T \sim 0.14$ keV and an energy that is $\sim 3\%$ of
the entire X-ray emission. To check the consistency, we fit the time dependent
BAT+XRT joint spectra by a Cutoff Power-law model (with absorption from both
Milk Way and the host galaxy, wabs*zwabs*cutoffpl in Xspec 12) and a
Blackbody+Cutoff Power-law model. We find that both models can give equally
acceptable fits to the data. For example, for the slice 2 (from $T_{\rm
trig}+240$ s to $T_{\rm trig}+734$ s), the former model gives $\chi^2/dof
=1305/1085$ with $N_{\rm gal}=7.05\times 10^{20}~{\rm cm^{-2}}$ and $N_{\rm
H,host}=4.6\times 10^{21} {\rm cm}^{-2}$, while the latter gives $\chi^2/dof
=1264/1085$ with a larger host galaxy absorption $N_{\rm H,host}= 1.3\times
10^{22} {\rm cm^{-2}}$ and a thermal temperature that is consistent with
\citet{Starling10}. Note that the peak of the proposed black body component is
near the low end of the XRT energy band and the neutral hydrogen absorption
around the thermal peak is large. This makes great uncertainty on
identification of such a thermal emission component from the data. Our spectral
analysis cannot robustly identify this component, although the $\chi^2$ is
slightly improved by adding it. Therefore, we do not claim a
thermal emission in the observed spectrum and only stick to the Cutoff
power-law model to discuss possible theoretical implications.
The second difference between the two events lies in
the supernova data. The modeling of SN 2006aj suggests a kinetic energy $\sim
2.5\times 10^{51}$ erg \citep{Mazzali06b}, much smaller than that of other SNe
associated with nearby GRBs (see Fig.\ref{fig:SN-GRB}). Although the modeling
of SN 2010bh is not available yet, current data imply an SN event as energetic
as SN 1998bw \citep{Chornock10}\footnote{With the simplest assumptions that (1) the opacity of the SN outflow is from Thompson scattering of
electrons and is $\sim 0.2~{\rm g^{-1}~cm^{2}}$ and (2) The density of expanding SN material takes the form $\propto R^{-k}$ \citep[$k\sim 6-8$, J. S. Deng, private communication, see also][]{MM99,BKC02,CF06}, it is straightforward to show that the mass of the SN material moving faster than $V_{\rm s}$ is $M_{\rm SN}(>V_{\rm s}) \sim 8\pi {k-1\over k-3}{m_p \over \sigma_{\rm T}}V_{\rm s}^{2}t^{2}$, where $V_{\rm s}$ is the photospheric velocity at $t$, $\sigma_{\rm T}$ is the Thompson scattering cross section and $m_{\rm p}$ is the rest mass of proton. The corresponding kinetic energy is $E_{\rm k,SN}(>V_{\rm s}) \sim 4\pi {k-1\over k-5}{m_p \over \sigma_{\rm T}}V_{\rm s}^{4}t^{2}$.
For SN 2010bh, $V_{\rm s}\sim 0.09 c$ at $t\sim 21$ day, we have $E_{\rm k,SN 10bh}(>0.09c)\sim 1.6\times 10^{52}~[(k-1)/3(k-5)]~{\rm erg}$. \label{foot:2}}, which is about one order of magnitude more
energetic than SN 2006aj.
So from the observational point of view, XRF 100316D/SN 2010bh and XRF
060218/SN 2006aj are not strictly ``twins''.

\section{A Possible Model for the long-lasting X-ray plateau}\label{sec:Interpretation}
In the following, we define $t = T-T_{\rm trig}+500$ s, i.e. the time elapse
since $T_{\rm trig}-500$ s. We interpret all the BAT/XRT data of XRF 100316D
for $0\leq t\leq 1.23\times 10^{3}~{\rm s}$ as ``prompt emission'' (i.e. the
radiation powered by some internal energy dissipation processes) for the
following two reasons. First, the steady plateau behavior observed in both BAT
and XRT band at $t \leq 1.23\times 10^{3}~{\rm s}$ with an evolving $E_p$ is
difficult to interpret within afterglow models. Second,
the sharp decline of the X-ray emission ($t^{-2}$ or even
steeper) expected in the time interval $1.23\times 10^{3}~{\rm s} < t <
3\times 10^{4}~{\rm s}$ resembles the early rapid decline that has been
detected in a considerable fraction of {\it Swift} GRBs, which is widely
taken as a piece of evidence of the end of prompt emission \citep{Zhang06}.
The nature of the X-ray emission detected at $t>3\times 10^{4}$ s is
hard to pin down. Its spectrum is very soft (photon index
$\Gamma=3.3^{+2.2}_{1.6}$), similar to that of XRF
060218. This is also unexpected in the external forward shock models,
and this late X-ray component may be related to a late central engine
afterglow, whose origin is unclear
\citep[e.g.,][]{Fan06,Soderberg06}.

The prompt BAT/XRT data do not show a significant variability
(Fig.\ref{fig:XRT}). The time-averaged $\gamma-$ray luminosity is
$\sim 3\times 10^{46}~{\rm erg~s^{-1}}$ and the X-ray luminosity is
$\sim 2\times 10^{46}~{\rm erg~s^{-1}}$. The bolometric luminosity of
the XRF outflow is therefore expected to be in the order of $10^{47}~{\rm
erg~s^{-1}}$. The duration of the BAT emission is at least $1.23
\times 10^3$ s, and can be longer. The relatively steady energy
output is naturally produced if {\it the central engine is a
neutron star with significant dipole radiation}. The dipole
radiation luminosity of a magnetized neutron star can be described as
\begin{eqnarray}
L_{\rm dip} = {2.6\times 10^{48}}~{\rm
erg~s^{-1}}~B_{\rm p,14}^2R_{\rm s,6}^6\Omega_4^4
\left(1+{t\over \tau_{\rm 0}}\right)^{-2},
\label{eq:E_inj}
\end{eqnarray}
where $B_{\rm p}$ is the dipole magnetic field strength of the neutron
star at the magnetic pole, $R_{\rm s}$ is the radius of the neutron
star, $\Omega$ is the angular frequency of radiation at $t=0$,
$\tau_{\rm 0}=1.6\times 10^4 B_{\rm p,14}^{-2} \Omega_4^{-2}I_{45}
R_{\rm s,6}^{-6}$ s is the corresponding spin-down timescale of the
magnetar, and $I\sim 10^{45}~{\rm g~cm^2}$ is the typical moment of
inertia of the magnetar (Pacini 1967; Gunn \& Ostriker 1969).
Here the convention $Q_{\rm n}= Q/10^{\rm n}$ is adopted in cgs units.
One then has $L_{\rm dip} \sim {\rm
const}~{\rm for} ~ t\ll \tau_{\rm 0}$ and $L_{\rm dip} \propto
t^{-2}~{\rm for} ~t\gg \tau_{\rm 0}$.  An abrupt drop in the
X-ray flux with a slope steeper than $t^{-2}$ may be interpreted
as a decrease of radiation efficiency, or the collapse of the
neutron star into a black hole,
possibly by losing the angular momentum or by accreting
materials.
Within such a model, the fact that
\[L_{\rm dip} \sim 10^{47}~{\rm erg~s^{-1}}, ~~~ \tau_{\rm 0} \sim 1000~{\rm s},\]
would require $(B_{\rm p,14},~\Omega_4,~I_{45},~R_{\rm s,6}) \sim
(30,~0.06,~1,~1)$.
This is a slow ($P \simeq 10$ ms) magnetar ($B_{\rm p} \simeq 3\times 10^{15}$ G).

\begin{figure}
 \includegraphics[angle=0,width=0.45\textwidth]{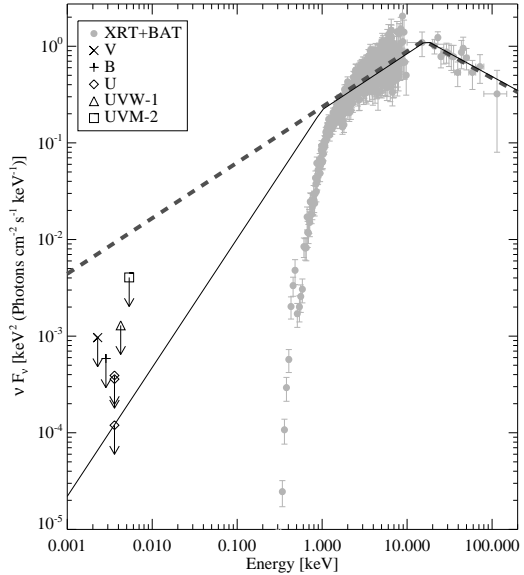}
\caption{Broadband SED from UVOT, XRT, and BAT data. Grey points show the time-averaged BAT+XRT spectrum between $T_{\rm trig}$+150 sand $T_{\rm trig}$+744 s. The thick dashed line represents an absorbed broken power-law fit (i.e., wabs*zwabs*bknpower in XSPEC although the absorption components are not plotted here) to the BAT+XRT data leading to $\Gamma_1=1.42$, $\Gamma_2=2.48$ and $E_{\rm break}=16~{\rm keV}$. The solid line shows the same fitting as above for higher energy band but with an additional break at 1 keV, below which photon index is set to $-2/3$. UVOT observations are taken from Starling et al. (2010).
 The extinction in each filter has been corrected by adopting $E_{\rm MW}(B-V)=0.12$ from the Milky Way and $E_{\rm host}(B-V)=0.1$ from the host galaxy (Starling et al.2010; Chornock et al. 2010) and a Milky Way-like extinction curve forall bands (Pei 1992). }
\label{fig:SED}
\end{figure}

The composition of this spindown-powered outflow is likely
Poynting-flux-dominated. Besides the magnetar argument (which
naturally gives a highly magnetized outflow), another argument
would be the lack of a bright
thermal component with a temperature
$kT \sim 10~{\rm keV}~L_{47}^{1/4}R_{0,9}^{-1/2}$ from the
outflow photosphere as predicted in the baryonic outflow
model, where $R_0$ is the initial radius where the outflow
is accelerated \citep[e.g.][]{ZP09,Fan10}. One may
argue that the photosphere radiation peaks at the observed $E_{\rm p}$.
Such a scenario, however, is hard to account for the X-ray
spectrum $F_\nu \propto \nu^{-0.32\pm 0.03}$.

Below we discuss a possible magnetic dissipation scenario that would
interpret XRF 100316D prompt emission.
By comparing the pair density ($\propto r^{-2}$, $r$ is the radial
distance from the central source) and the density required for
co-rotation ($\propto r^{-1}$ beyond the light cylinder of the compact
object), one can estimate the radius at which the MHD condition breaks
down, which reads $r_{\rm MHD} \sim 5\times 10^{15} L_{47}^{1/2}
\sigma_{1.6}^{-1}t_{v,m,-2} \Gamma_{\rm i,1.5}^{-1} ~{\rm cm}$, where
$\sigma$ is the ratio of the magnetic energy flux to the particle
energy flux, $\Gamma_{\rm i}$ is the bulk Lorentz factor of the
outflow, $t_{v,m} \sim P$ is the minimum variability timescale of the
central engine \citep{ZM02,Fan05,Gao06}. Beyond this radius, intense
electromagnetic waves are generated and outflowing particles are
accelerated \citep{Usov94,Lyutikov01}.
Part of the Poynting flux energy is converted to radiation.
At $r_{\rm MHD}$, the comoving magnetic fields $B_{\rm
MHD}$ can be estimated as $B_{\rm MHD}\sim 20~ \xi \sigma_{1.6}
t_{v,m,-2}^{-1}~{\rm G}$, where $\xi\leq 1$ reflects the
efficiency of
magnetic energy dissipation.  When magnetic dissipation occurs, a
fraction $\varepsilon_{\rm e}$ of the dissipated comoving magnetic
energy would be eventually converted to the comoving kinetic energy of
the electrons. Electrons may be linearly accelerated in the electric
fields or stochastically accelerated in the random electromagnetic
fields. One may assume that the accelerated electrons form a single
power-law distribution in energy, i.e. $dn/d\gamma_{\rm e} \propto
\gamma_{\rm e}^{\rm -p}$ for $\gamma_{\rm e}>\gamma_{\rm e,m}$,
where $\gamma_{\rm e,m}$ can be estimated as $\gamma_{\rm e,m}\sim
3.5\times 10^{4}~\sigma_{1.6} C_p$, and $C_p \equiv
(\varepsilon_e/0.5)(p-2)/(p-1)$.
At $r_{\rm MHD}$, the corresponding synchrotron radiation frequency is
\citep{Fan05}
\begin{equation}
\nu_{\rm m,MHD}\sim 2\times 10^{18}~{\rm Hz}~\xi \sigma_{1.6}^3
C_p^2 \Gamma_{\rm i,1.5}t_{v,m,-2}^{-1}.
\label{eq:nu_m}
\end{equation}
The cooling Lorentz factor can be estimated by $\gamma_{\rm e,c}\sim
4.5\times 10^{19}\Gamma_{\rm i} /(r_{\rm MHD} B_{\rm MHD}^2)$, which
gives a synchrotron cooling frequency
$\nu_{\rm c,MHD}\sim 6\times 10^{14}~{\rm Hz}~\Gamma_{\rm
i,1.5}^{5}L_{47}^{-1} \xi^{-3}\sigma_{1.6}^{-1}t_{\rm v,m,-2}$.
The observed XRT spectrum can be approximated by $F_\nu \propto
\nu^{-0.3\sim-0.4}$, which is close to the fast cooling spectrum
$\nu^{-1/2}$, suggesting $\nu_{\rm c,MHD}\lesssim
 10^{17}~{\rm Hz}$ (it is straightforward to show that for the fiducial parameters adopted in this work the synchrotron self-absorption is
 below the optical band).
On the other hand, the non-detection of prompt
emission by UVOT \citep{Starling10} favors a high cooling
frequency $\nu_{\rm c,MHD} \sim 2.4\times 10^{17}$ Hz. As shown in Fig.\ref{fig:SED},
the synchrotron radiation model with such a high $\nu_{\rm c,MHD}$ can roughly account for the data.
We then get a constraint
\[\Gamma_{\rm i} \sim 60
L_{47}^{1/5}\xi_{-0.3}^{3/5}\sigma_{1.6}^{1/5}t_{\rm
v,m,-2}^{-1/5}(\nu_{\rm c,MHD}/2.4\times 10^{17}~{\rm Hz})^{1/5}.\]
Substituting this into
eq.(\ref{eq:nu_m}) we get
\[\sigma \sim 54 L_{47}^{-1/16}({\nu_{\rm c,MHD}\over
2.4\times 10^{17}~{\rm Hz}})^{-1/16}\xi_{-0.3}^{-1/2}C_{\rm p}^{-5/8}
t_{\rm v,m,-2}^{3/8}.\]
Within such a scenario, $E_{\rm p}$ is defined by $\nu_{\rm m,MHD}$
[Eq.(\ref{eq:nu_m})]. An observed evolving $E_{\rm p}$ may be understood
by invoking a decreasing $\sigma$, which is consistent with the
magnetic field dissipation hypothesis. For $\nu>E_{\rm p}/h$ (where $h$ is Planck's constant), the spectrum
is $F_\nu \propto \nu^{-p/2}$.

The typical variability timescale of the radiation powered by the
magnetic energy dissipation can be estimated as
\begin{equation}
\delta t \sim r_{\rm MHD}/(2\Gamma_{\rm i}^2 c) \sim 10~{\rm s}~L_{47}^{1/2}
(\sigma/50)^{-1}t_{v,m,-2} (\Gamma_{\rm i}/60)^{-3},
\end{equation}
which is much larger than the rotation period $P \sim 10~{\rm ms}$ of
the magnetar, implying that the pulsation of the central engine is
undetectable.

\section{Discussion and Speculations}
So far the nearby supernova-associated GRBs, except GRB 030329, are found to
be intrinsically under-luminous. They share the similarities
such as low isotropic energies and smooth light curves,
but differ in some aspects. For example, GRB 980425 and GRB 031203
have shorter durations and higher $E_{\rm p}$'s than
XRF 060218 and XRF 100316D. The underlying physical processes
that result in these differences are not
well understood.
Through supernova modeling, it is found that GRB 980425
and GRB 031203 have a progenitor star massive enough to form
a black hole \citep{Deng05,Mazzali06a},
while XRF 060218 has a less massive progenitor that most
plausibly produces a neutron star \citep{Mazzali06b}.
The luminosity and the duration of XRF 100316D are consistent with the
radiation from a neutron star with a dipole magnetic field $B_{\rm p}
\sim 3\times 10^{15}$ G and a rotation period $P \sim 10$ ms.
This seems to point towards a hypothesis that two types of central
engines define the apparent dichotomy of the SN-associated LL-GRBs,
i.e. black hole engines give rise to ``shorter" and ``harder" GRBs such as GRB 980425 and GRB 031203, while
magnetar engines give rise to very long and soft XRFs such as XRF 060218 and XRF 100316D \footnote{We
caution that such a scenario is not robust, since
a magnetar engine may also drive SN 1998bw \citep{Woosley09},
and a black hole engine may be also able to reproduce the XRF
100316D-like light curves through tuning the parameters of fall-back
materials and arguing for a Poynting-flux-dominated outflow from
a highly magnetized black hole engine
\citep[e.g.][]{MacFadyen01,ZhangW08}.}.

Adding in SN data makes the scenario more complicated.
Although SN 2006aj associated with XRF 060218 does not conflict with
a slow magnetar central engine, SN 2010bh associated with XRF 100316D
may be too energetic to be interpreted with a slow magnetar central
engine. If it is confirmed that the kinetic energy of SN 2010bh is
in excess of $10^{52}$ erg, neither the neutrino energy nor the
magnetar spin energy ($\sim 10^{50}$ erg) are adequate to power
the SN. A salient feature of the dipole spindown formula
(Eq.[\ref{eq:E_inj}]) is that if one shifts the time zero point
(e.g. $t'=t - t_0$), the spindown law still applies, with the initial
angular frequency re-defined as $\Omega'=\Omega(t'=0)$, and the
characteristic spindown time scale re-defined as
$\tau'_0=1.6\times 10^4 B_{\rm p,14}^{-2} {\Omega'_4}^{-2} I_{45}
R_{s,6}^{-6}$. This suggests that the observed plateau feature can
be still interpreted if the initial period is much shorter than
10 ms, say, $P_0 \sim 1$ ms, if the time zero point is much earlier
than $T_{\rm trig}-500$ s (i.e. $t=0$).
This is because a power-law decay light curve
may show an artificial plateau in the log-log space, if the zero
time is mis-placed to a later epoch \citep{Yamazaki09,Liang09}.
Within such a scenario, a nascent magnetar was born with an initial
period $P_0 \sim 1$ ms at $t \sim -5\times 10^3$ s.
Its intial dipole radiation
was trapped by the envelope of the progenitor and could not escape.
This spindown energy gives enough impetus to explode the
star and power the energetic SN 2010bh. After a significant delay
($\sim 5\times 10^3$ s to spin down from 1 ms to 10 ms for $B_{\rm p}
\sim 3\times 10^{15}$ G), the magnetar wind finally
managed to escape as a relativistic Poynting-flux-dominated outflow.
An observer noticed the jet emission only around $t=0$. The above argument also
applies to the model of fallback accretion onto a nascent black hole.

With such a hypothesis, one may envision a unified picture to
understand the diversity of GRB/SN associations, by invoking
a variety of initial powers and the delay
times between the core collapse and the emergence of the
relativistic jet from the star. The speculation is the following:
\begin{itemize}
\item To produce an energetic SN/luminous GRB
(e.g. GRB 030329/SN 2003dh), the central engine is powerful (a black
hole with an accretion disk or a rapidly spinning magnetar) and the relativistic outflow
can break out the progenitor soon enough when the engine is still
working effectively.
\item To produce an energetic SN / underluminous GRB
(e.g. GRB 980425/SN 1998bw, GRB 031203/SN 2003lw, and
XRF 100316D/SN 2010bh), the central engine is initially powerful,
but it takes time for the relativistic wind to emerge from the
star. As it breaks out the star, the central engine already
fades down with a decreased power. The longer, softer XRFs are
probably powered by a magnetar, while the shorter, harder
GRBs are probably powered by a black hole.
\item To produce a less-energetic SN / underluminous GRB
(e.g. XRF 060218/SN 2006aj),
the central engine is a slow magnetar with an initial rotation
energy less than $10^{51}$ ergs. The emergence of the relativistic
outflow can be prompt or somewhat (but not significantly) delayed.
\end{itemize}

Finally, a straightforward expectation from the speculation that
XRF 100316D outflow is Poynting-flux-dominated
is that the prompt emission should be linearly polarized
\citep[e.g.,][]{Fan05}.
The polarimetry measurements of events such as XRF 100316D and XRF
060218 would provide a criterion to differentiate this model
from the shock breakout model, which does not predict a strong
polarization signal.

\section*{Acknowledgments}
We thank the anonymous referee for helpful comments, and S. Covino, J. S. Deng and  R. L. C. Starling for communications. This work was supported in part by
the National basic research program of China under grant
2009CB824800 (for Y.Z.F. and E.W.L), and by NASA NNX09AT66G, NNX10AD48G, and NSF AST-0908362 (for B.Z.).

\clearpage

\end{document}